\documentclass[12pt]{article}
\usepackage{latexsym}
\usepackage{amssymb}
\usepackage{amsmath}
\usepackage{graphicx}
\usepackage{cite}

\begin{document}
\title{Quasi-eikonal and quasi-U-matrix unitarization schemes beyond the Black Disk Limit}
\author{E. Martynov \thanks{Bogolyubov Institute for Theoretical Physics,
              Metrolologichna 14b, Kiev, UA-03680, Ukraine} \thanks{email: martynov@bitp.kiev.ua}}
\maketitle
\begin{abstract}
Quasi-eikonal and quasi-U-matrix unitarization of the standard Regge-pole amplitude for $\alpha (0)>1$ have been considered. We show that some violation of unitarity even at high energy exists in both models. We have found in quasi-eikonal model a bump-oscillation structure of ${\rm Im}H(s,b)$ at large values of impact parameter $b$ but where ${\rm Im}H(s,b)$ is closed to the maximal value. We argue that it is possible to choose the parameter regulating deviation of generalized models from pure eikonal or U-matrix modes in order to restore unitarity.
\end{abstract}

\begin{figure}[h*]
  \centering
  \includegraphics[scale=0.3]{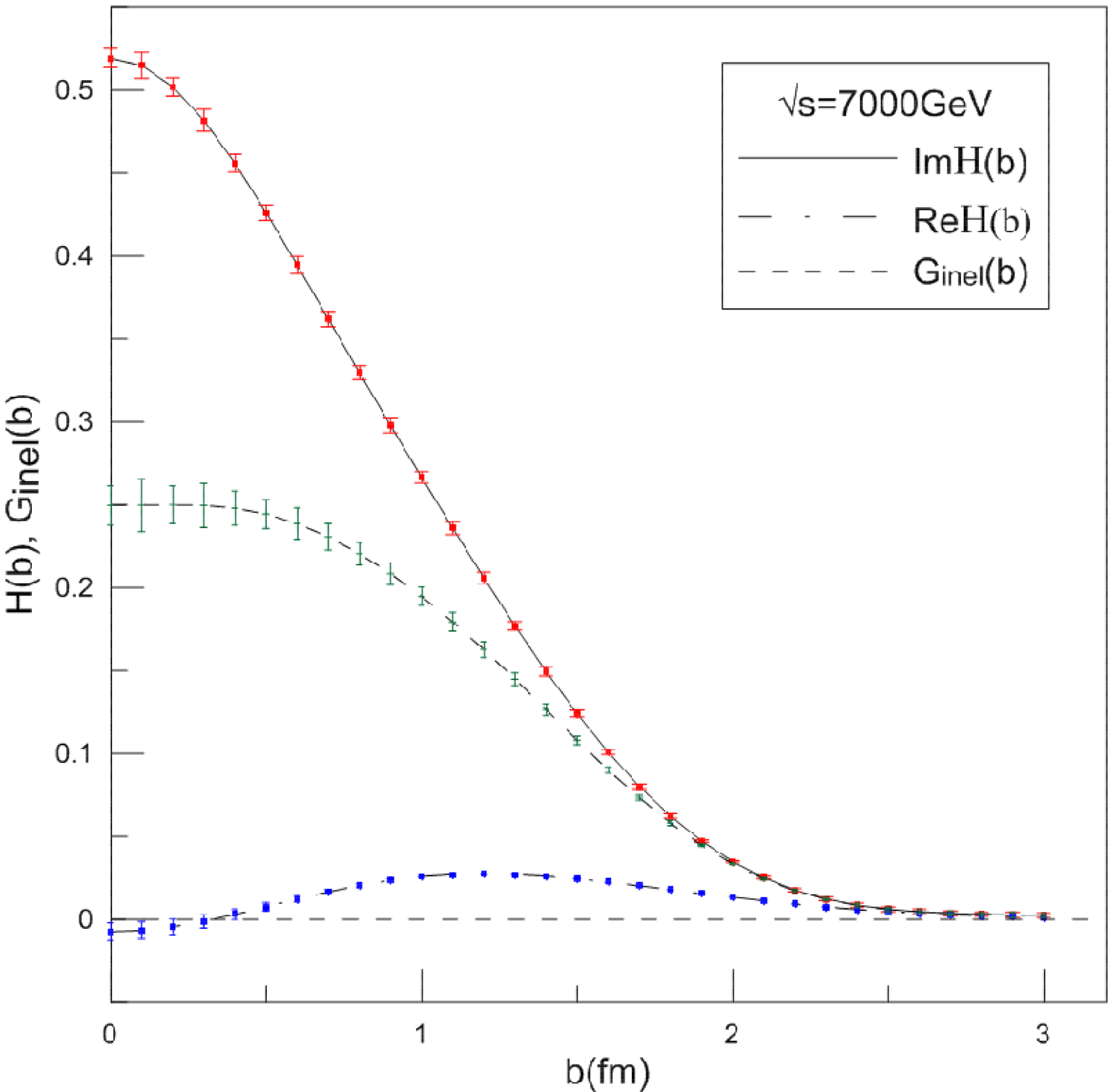}\\
  \caption{``Experimental''  data for $H(s,b)$ extracted from the TOTEM $d\sigma/dt$ data at $\sqrt{s}=$7 TeV }\label{fig:ImHdata}
\end{figure}

It was shown in the recent paper \cite{AKMT} that the impact-parameter amplitude $H(s,b)$ extracted from  $pp$ elastic scattering data of the TOTEM experiment at $\sqrt{s}=7$ TeV \cite{totem} exceeds the black disk (BDL) limit ${\rm Im}H(s,0)=1/2$.
We define $H(s,b)$ as the following  transformation of standard scattering amplitude (at high $s$)
\begin{equation}\label{eq:defH}
  H(s,b)=\frac{1}{8\pi s}\int\limits_{0}^{\infty}dq\,qJ_{0}(qb)A(s,t=-\vec{q}^{2}), \qquad \sigma_{t}=\frac{1}{s}{\rm Im} A(s,0), \quad \frac{d\sigma }{dt}=\frac{1}{16\pi s^{2}}|A(s,t)|^{2}.
\end{equation}
The extracted data for ${\rm Im}H(s,b)$ at $\sqrt{s}=7$ TeV  are shown in Fig. \ref{fig:ImHdata} (the figure is taken from the \cite{AKMT}).

\begin{figure}[h*]
  \centering
  \includegraphics[scale=0.45]{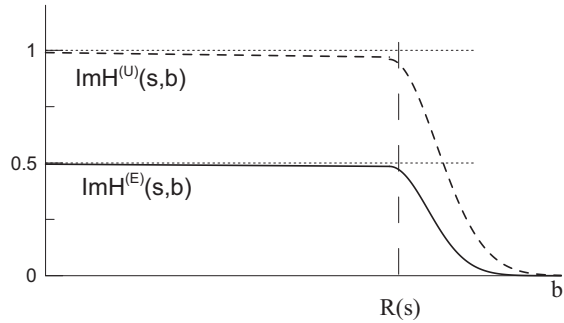}\\
  \caption{Elastic impact-parameter amplitude unitarized by eikonal or U-matrix method }\label{fig:ImH}
\end{figure}

This result, provided that it will be confirmed at higher energies (8, 13, 14 TeV at LHC), leads to the important consequences for many phenomenological models constructed within a hypothesis that the BDL regime is realized in hadron elastic scattering at high energy. First of all it concerns with a widely explored eikonal model
\begin{equation}\label{eq:eikonal}
 2iH^{(E)}(s,b)=e^{2ih(s,b)}-1
\end{equation}
where usually and in accordance with Regge approach an input amplitudes $a(s,t)$ and $h(s,b)$ are assumed to have the following properties.
\begin{itemize}
\item Amplitude $a(s,t)$ is presumably imaginary at least at small $t$.
\item Amplitude $h(s,b)\propto i(-is/s_{0})^{\varepsilon }$ where $\varepsilon>0$ and $s_{0}=1$ GeV at fixed impact parameter $b$ and at $s\to \infty$.
\item Amplitude $h(s,b)\propto \exp(-b^{2}/4R_{0}^{2}(s))$ at $b\gg R_{1}(s)$ where $R_{0}^{2}(s)\approx \alpha '\ln(s/s_{0})$ and $R_{1}(s)\propto \ln(s/s_{0})$. \footnote{However, it is known \cite{Gribov} that in this region (where $H(s,b)\approx h(s,b)$) the impact-parameter amplitude must have an exponential, $H(s,b)\propto \exp(-bm)$, where $m$ is a constant, rather than gauss  behavior caused by contribution of Regge pole with linear trajectory.}
\end{itemize}
Then it can be concluded that the unitarized amplitude $H^{(E)}(s,b)$ at $s\to \infty$ is coming close to 1/2 at $b\lesssim R_{1}(s)$ and going to zero at $b\gg R_{1}(s)$ as shown in Fig. \ref{fig:ImH}.

In another unitarization method, $U$-matrix unitarization \cite{TroshinTyurin}, the output amplitude $H^{(U)}(s,b)$ is expressed through $h(s,b)$ by the following equation
\begin{equation}\label{eq:U-m}
H^{(U)}(s,b)=\frac{h(s,b)}{1-ih(s,b)}.
\end{equation}
At $b\lesssim R(s)$ the impact-parameter amplitude ${\rm Im}H^{(U)}(s,b)\to 1$ because ${\rm Im}h(s,b)\to \infty$. The height of a step in $H(s,b)$ is not equal to 1 at any value of $s$ but it goes to 1 when $s\to \infty$.

Thus the $U$-matrix unitarization does not lead to any contradiction with  above mentioned impact-parameter analysis of the TOTEM data  while in the eikonal method the maximal value of amplitude ${\rm Im}H(s,b)=1/2$ conflicts with the above mentioned analysis of the TOTEM data.

Is it possible to fix the problem in eikonal or eikonal-type approach? Several modifications and generalizations of the eikonal as well as of the U-matrix methods have been suggested and explored long ago and recently \cite{TroshinTyurin, TerM, GKMS, DGMP, CPS, SCP, Kancheli, ANN}. In what follows we will concentrate on two models, quasi-eikonal \cite{TerM} and quasi-$U$-matrix. They possess some interest and importance because a choice of additional parameter in these models allows to obtain asymptotical regime with the maximal value of the impact amplitude step between 1/2 and 1, i.e $1/2<{\rm Im}H(s,b)\leq 1$.

It is worthwhile to remind that the both models can be treated on the unique basis by summing multireggeon exchanges in $s$-channel or multiple rescatterings. Namely, elastic scattering amplitude at high $s$ can be written as sum of  $n$-Pomeron exchanges.
\begin{equation}\label{eq:rescat}
H(s,b)=\frac{1}{2i}\sum_{n=1}^{\infty}\frac{\left (N^{(n)}(b)\right )^{2}}{n!}(2ih(s,b))^{n}
\end{equation}
where
\begin{equation}\label{eq:inputP}
\begin{array}{rl}
h(s,b)=&\displaystyle ig(-is/s_{0})^{\varepsilon }\frac{\exp({-b^{2}/4R^{2}(s)})}{2R^{2}(s)},\\
g=& {\rm const},\quad s_{0}=1 {\rm GeV}^{2},\quad \varepsilon =\alpha (0)-1,\\
R^{2}(s)=&\beta +\alpha '\ln(s/s_{0}), \quad \beta ={\rm const}.
\end{array}
\end{equation}
The function $N^{(n)}(b)$ is the vertex function that describes an interaction of two hadrons with $n$ pomerons, each of them being characterized by the impact parameter $\vec{b}$. The simplest assumption is that $N^{(n)}(b)$ depends only on number of pomerons, $n$. Then we define $\lambda (n)=(N^{(n)}(b))^{2}$ and obtain
\begin{equation}\label{eq:rescat-lambda}
H(s,b)=\frac{1}{2i}\sum_{n=1}^{\infty}\frac{\lambda (n)}{n!}(2ih(s,b))^{n}.
\end{equation}
Let us consider two explicit forms of $\lambda (n)$.

 Assuming the first form  $\lambda (n)=\lambda _{E}^{n-1}$ one can immediately have the quasi-eikonal (QE) model \cite{TerM} (in the pure eikonal model $\lambda _{E}$ equals to 1)
\begin{equation}\label{eq:QE}
 H(s,b)=\frac{e^{2i\lambda _{E} h(s,b)}-1}{2i\lambda _{E}}.
\end{equation}

The second assumption that $\lambda (n)=n!\lambda _{U}^{n-1}$ leads to quasi-$U$-matrix (QU) model (in the original pure $U$-matrix model \cite{TroTyu-1} $\lambda _{U}$ equals to 1/2).  Then
\begin{equation}\label{eq:QU}
 H(s,b)=\frac{h(s,b)}{1-2i\lambda _{U} h(s,b)}.
\end{equation}

It follows from Eqs.~(\ref{eq:QE},\ref{eq:QU}) that $|H(s,b)|\to 1/(2\lambda _{E,U}$) at $s\to \infty$ and $b<R_{1}(s)=2\sqrt{\alpha '/\varepsilon }\ln(s/s_{0})$. Then from the unitarity inequality $|H(s,b)|\leq 1$ we have  $\lambda _{E,U}\geq 1/2$. Thus, if $1/2\leq \lambda _{E,U} \leq 1$ then $1/2\leq {\rm Im}H(s,b)\leq 1$ at $s\to \infty$ in the region $b< R_{1}(s)$.

Many pro and contra arguments concerning QE-model were discussed in \cite{CPS, SCP, TroshinTyurin, Kancheli,Troshin}. In the present paper we give  some additional arguments that such a model can be consistent with unitarity restrictions on $H(s,b)$. We would like to notice that the result of unitarization depends on both ingredients: input amplitude $a(s,t)$ (or $h(s,b)$) and scheme that determines a dependence of output amplitude $H(s,b)$ on input $h(s,b)$. We would like to notice that it is not necessary to require (as is mentioned in \cite{Troshin}) that the whole upper semi-plane (${\rm Re}h,{\rm Im}h$) must be mapped under unitarization to the unitarity circle in (${\rm Re}H,{\rm ImH}$). Such a requirement can be valid only for a part of the semi-plane (${\rm Re}h,{\rm Im}h$) depending on the specific model for input $h(s,b)$.

Any scheme does not guaranty the correct properties of $H(s,b)$ for an arbitrary $h(s,b))$. However, it is necessary for output amplitude to satisfy the unitarity requirements. From the unitarity equation
\begin{equation}\label{eq:unitarity}
{\rm Im}H(s,b)=|H(s,b)|^{2}+G_{inel}(s,b)>0
\end{equation}
where $G_{inel}(s,b)$ takes into account a contribution of inelastic processes, which is positive at $s>9m^{2}$ (for the sake of simplicity we consider identical hadrons with mass $m$). Unitarity imposes some restrictions on the both ingredients $H(s,b)$ and $h(s,b)$ of the unitarization scheme. It follows from a positivity of $G_{inel}$ that
\begin{equation}\label{eq:unitar-ineq}
{\rm Im}H(s,b)-|H(s,b)|^{2}>0.
\end{equation}

Let us consider now a quasi-eikonal model (\ref{eq:QE}) with $1/2\leq \lambda \leq 1$ and take a contribution of simple Regge pole defined in Eq. (\ref{eq:inputP}) as input amplitude.  Inequality $\lambda >1$ makes no sense because in this case ${\rm Im}H(s,b)<1/2$ at $s\to \infty$, that is not supported by data. Of course, other contributions ($f, \omega $ reggeons, for example) are important at not very high energy, but at energies $\sqrt{s}$ exceeding or approximately equal to few hundred GeV they are very small and can be neglected in the given qualitative analyses. Let rewrite $H(s,b)$ as follows
\begin{equation}
H(s,b)=-\frac{i}{2\lambda }\left [ e^{2i\lambda (h_{r}+ih_{i})}-1\right ]=-i(-1+\mu C+i\mu S)/2\lambda 
\end{equation}
where $h_{i}={\rm Im}h(s,b), h_{r}={\rm Re}h(s,b), \mu =\exp(-2\lambda h_{i}),\quad  C=\cos(2\lambda h_{r}), \quad S=\sin(2\lambda h_{r})$.
The inequality (\ref{eq:unitar-ineq}) in terms of $\mu $ and $C$ has the form
\begin{equation}\label{eq:ineq-0}
2\lambda (1-\mu C)\geq \mu ^{2}-2\mu C+1, \qquad \mu <1, \qquad |C|\leq 1.
\end{equation}

1. If $b< R_{1}(s)=2\sqrt{\alpha '/\varepsilon }\ln(s/s_{0})$ and $s\gg s_{0}$ then $|h(s,b)|\gg 1$ and $\mu \ll 1$, inequality (\ref{eq:ineq-0}) is read as $\lambda \geq 1/2$.

2. If $b\gg R_{1}(s)$ then $|h(s,b)|\ll 1$ and $H(s,b)\approx h(s,b)$ independently of energy and $\lambda $. Obviously that in this case ${\rm Im}H(s,b)\gg |H(s,b)|^{2}$ at any $\lambda $ if ${\rm Im}h(s,b)>0$.

One can show that the above two conclusions 1. and 2. are still valid for the QU-model (\ref{eq:QU}) as well.

\begin{figure}[h*]
 \centering
\includegraphics[scale=0.6]{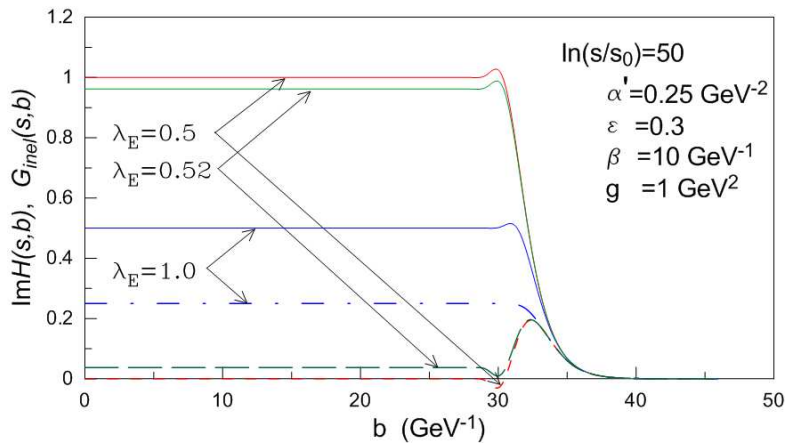}
  \caption{Restoring unitarity bounds on ${\rm Im}H(s,b)$ and $G_{inel}(s,b)$ by changing $\lambda _{E}$ in QE-model}\label{fig:H-EvsQE}
\end{figure}

3. It is impossible to analyze analytically the unitarity inequalities at nonasymptotic energy and in a region of $b$ where $H(s,b)$ is rapidly decreasing. That is why we calculate numerically ${\rm Im}H(s,b)$ and $G_{inel}(s,b)$ at some typical parameters of input amplitude $h(s,b)$,  Eq. (\ref{eq:inputP}).
To this end we have fixed parameters $g$ and $\alpha ', \beta $ (given in GeV$^{-2}$).  Their values and results obtained for ${\rm Im}H(s,b)$ and $G_{inel}(s,b)$ with various parameters $\varepsilon , \lambda $ as well as energy $\sqrt{s}$ (GeV) are shown in Figs. \ref{fig:H-violation},\ref{fig:H-E-U-s},\ref{fig:H(b-b0)},\ref{fig:H-EvsQE}.

\begin{itemize}
\item
We have found that unitarity inequalities (\ref{eq:unitarity}) and (\ref{eq:unitar-ineq}) are violated at ``low'' energy in both models (QE at $\lambda _{E}\geq 1/2$ and  QU at $\lambda _{U}\geq 1/2$). It is well pronounced if the bare pole intercept $\varepsilon > \approx 0.15-0.16$ in QE-model and $\varepsilon > \approx 0.35-0.4$ (see Fig. \ref{fig:H-violation}). It is not clear how to fix the problem. Maybe a nonlinear trajectory of pomeron or an additional (non Regge-pole?) term in $h(s,b)$ can help.
\item
The violation is saving in QE-model while it is absent in QU-model at increasing energy. (Fig. \ref{fig:H-E-U-s}). The violation is small but it exists. Let us consider a region $b\lesssim b_{0}=2\sqrt{\varepsilon Re( \ln(-is/s_{0})R^{2}(s))}$, where $b_{0}$ is determined from the following equation
\begin{equation}\label{eq:b0}
Re[\varepsilon \ln(-is/s_{0})-b_{0}^{2}/4R^{2}(s)]=0.
\end{equation}

\begin{figure}[h*]
 \centering
\includegraphics[scale=0.7]{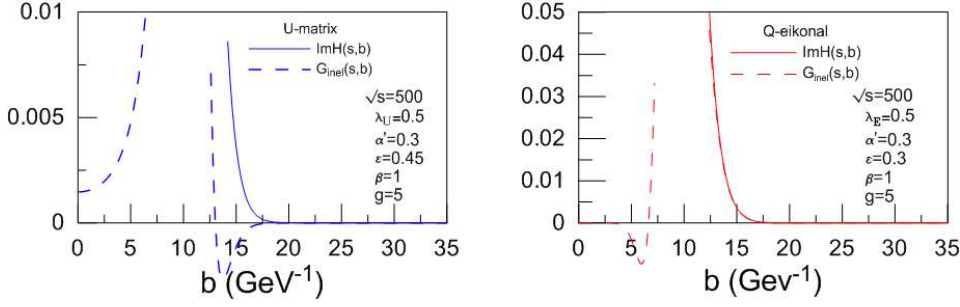}
  \caption{The explicit example of the unitarity violation for $G_{inel}(s,b)$ in QE- and U-models,}\label{fig:H-violation}
\end{figure}

\begin{figure}[h*]
 \centering
\includegraphics[scale=.5]{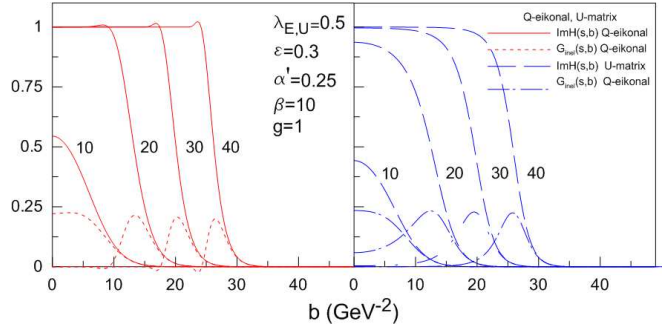}
  \caption{${\rm Im}H(s,b)$ and $G_{inel}(s,b)$ in QE-model (left) and U-model (right) at various energies (values of $\ln(s/s_{0})$ are indicated near the curves ${\rm Im}H(s,b)$ ). }\label{fig:H-E-U-s}
\end{figure}

\begin{figure}[h*]
 \centering
\includegraphics[scale=0.55]{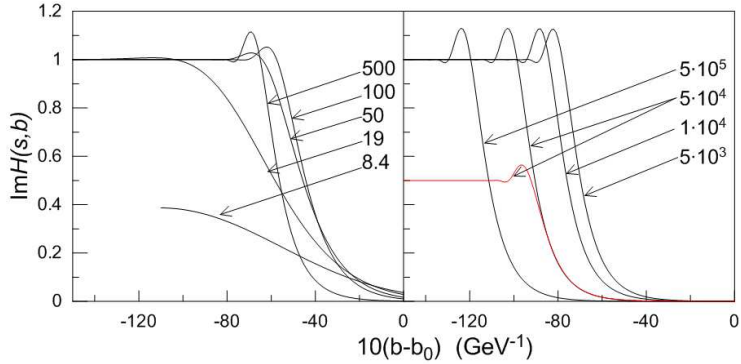}
  \caption{Bump effect in the QE-model ($\lambda _{E}=0.5$) at various energies with $\xi =\ln(s/s_{0})$,  $\xi $-values are given in Figure, corresponding curves are pointed by arrows.  $b_{0}=2\sqrt{\varepsilon Re( \ln(-is/s_{0})R^{2}(s))}$. Calculations were made with parameters: $\varepsilon =0.3, \alpha '=0.25 {\rm GeV}^{-2}, \beta =10{\rm GeV}^{-2}, g=1$. The red line illustrates ${\rm Im}H(s,b)$ in pure eikonal model ($\lambda _{E}=1$) at $\xi =50000.$ }\label{fig:H(b-b0)}
\end{figure}

\item
One can see (Fig. \ref{fig:H(b-b0)}) a bump followed by small oscillations (they can be seen well on the larger scale of the figure) above ${\rm Im}H(s,b)=1$ in QE-model at $\lambda _{E}=1/2$. It is important to notice that such a bump appears as well in the pure eikonal model with $\lambda _{E}=1$ but above ${\rm Im}H(s,b)=1/2$. It is illustrated in Fig. \ref{fig:H(b-b0)} on the right panel.  Thus we conclude that the black disk bound ${\rm Im}H(s,b)\leq 1/2)$ is violated in the pure eikonal unitarization of a simple pole with a linear trajectory and $\alpha (0)>1$. We would like to notice that the same oscillation-bump structure exists in another generalization of E- and U-models considered in \cite{CPS}
\begin{equation}\label{eq:interpol-QE-QU}
H(s,b)=\frac{i}{2\lambda }\left [1-\frac{1}{(1-2i\lambda h(s,b)/\gamma )^{\gamma }}\right].
\end{equation}
It interpolates between QU-model (at $\gamma =1$) and QE-model (at $\gamma \to \infty$). This structure appears at $\gamma \gtrsim 2$ if input amplitude $h(s,b)$ is the contribution of simple pomeron pole with line trajectory and $\alpha (0)>1$.
\item
The second important observation consists in a constant height of bump in a huge energy interval from ($\ln(s/s_{0}\sim 500$ up to $5\cdot10^{5}$) and most likely at higher $s$). It can be verified that bump height depends on $\varepsilon $ while it does not depend on energy at very high $s$. The same structure is appeared as well if pomeron trajectory is nonlinear (for example if $\alpha (t)=\alpha (0)+\gamma (\sqrt{t_{0}}-\sqrt{t_{0}-t})$). However there is no such a structure if a term $\propto \exp(-b/b_{0})$ (where $b_{0}$ is constant) is added to $h(s,b)$.
\item
It follows from these listed properties that unitarity violation in QE-model can be eliminated by an appropriate choice of $\lambda _{E}$, that is demonstrated in Fig. \ref{fig:H-EvsQE}. The minimal value of $\lambda _{E}$ is larger than 1/2 and depends on the input pomeron intercept $\varepsilon=\alpha (0)-1$.
\end{itemize}

\section{Conclusion and Aknowledgements}
Basing on the performed numerical analysis of the considered explicit examples for input amplitudes we can claim that
\begin{itemize}
\item
Quasieikonal unitarization model under suitable choice of $\lambda _{E}$ allows to describe scattering amplitude beyond the Black Disk Limit.
\item
Both the QE and QU unitarization schemes can describe a possible (hypothetic) regime of hadron interaction where $1/2<{\rm Im}H(s,b)<1$ at $s\to \infty$ and $b<const\ln(s/s_{0})$.
\item
It is important to check out carefully (at least numerically) how the unitarity bounds for $H(s,b)$ are satisfied when specific model for input amplitude $h(s,b)$ is taken.
\item
Input simple pomeron pole must be modified before unitarization is applied in order to avoid an oscillation of ${\rm Im}h(s,b)$ in a region where $h(s,b)\to 0$.
\end{itemize}

Author thanks S.M. Troshin for many challenging and fruitful discussions. The work is supported by the Department of Nuclear Physics and Power Engineering of the National Academy of Sciences of Ukraine (grant No CO-2-1/2014).

\end{document}